\documentclass{article}

\usepackage{PRIMEarxiv}

\usepackage[utf8]{inputenc} 
\usepackage[T1]{fontenc}    
\usepackage{hyperref}       
\usepackage{url}            
\usepackage{booktabs}       
\usepackage{amsfonts}       
\usepackage{nicefrac}       
\usepackage{microtype}      
\usepackage{lipsum}
\usepackage{fancyhdr}       
\usepackage{graphicx}       
\graphicspath{{media/}}     

\usepackage{amssymb}
\usepackage{amsmath}
\usepackage[table,xcdraw]{xcolor}
\usepackage{multirow}

\title{Multi-Period Portfolio Optimisation Using a Regime-Switching Predictive Framework
}

\author{
  Piotr Pomorski, Denise Gorse \\
  Department of Computer Science \\
  University College London \\
  London\\
  \texttt{\{P.Pomorski, D.Gorse\}@cs.ucl.ac.uk} \\
}

\begin{document}
\maketitle

\begin{abstract}
Regime-switching poses both problems and opportunities for portfolio managers. If a switch in the behaviour of the markets is not quickly detected it can be a source of loss, since previous trading positions may be inappropriate in the new regime. However, if a regime-switch can be detected quickly, and especially if it can be predicted ahead of time, these changes in market behaviour can instead be a source of substantial profit. The work of this paper builds on two previous works by the authors, the first of these dealing with regime detection and the second, which is an extension of the first, with regime prediction. Specifically, this work uses our previous regime-prediction model (KMRF) within a framework of multi-period portfolio optimisation, achieved by model predictive control, (MPC), with the KMRF-derived return estimates accuracy-boosted by means of a novel use of a Kalman filter. The resulting proposed model, which we term the \textit{KMRF+MPC model}, to reflect its constituent methodologies, is demonstrated to outperform industry-standard benchmarks, even though it is restricted, in order to be acceptable to the widest range of investors, to long-only positions.
\end{abstract}


\section{Introduction}
Regime shifts, typically between periods of low-volatility market growth and periods of high-volatility contractions, are a source of both risk and opportunity for portfolio managers. In order to either reduce losses or take advantage of a new opportunity for profit, it is necessary at minimum to be able to detect that a new regime has been entered. In~\cite{pomorski2023improving}, we presented a novel model for regime detection, KAMA+MSR, that combined statistical (Markov-switching regression (MSR)) and technical (Kaufman's adaptive moving average~\cite{kaufman1995smarter} (KAMA)) tools to effectively detect regime change and act upon it profitably. 
However, in this type of model there is inevitably a loss-generating lag, as the new regime cannot be detected and acted upon instantaneously.
In~\cite{pomorski2023improving} we built upon this work, using the KAMA+MSR model to generate regime labels to be predicted by a machine learning model. A set of three random forest (RF) models predicted ex-ante 'bullish' and 'bearish' regimes within equity, commodity, and FX asset classes, achieving not only high classification scores out of sample, but also high Sortino ratios, indicating solid financial performance when the models were applied to trading.
However, despite its excellent trading results, this model, which we termed the KMRF model (KAMA+MSR+RF), had three limitations: first, that its use of frequent shorting positions could be problematic for the many institutions disallowed from using these; second, that it used only fixed, equal, weightings for the assets; and third, that it assumed fixed transaction costs, independently of the liquidity and volatility of the traded assets.

All three of these limitations will be addressed by the portfolio optimisation method presented in the current work. This method will enhance the KMRF model via the use of \textit{model predictive control} (MPC)~\cite{samuelson1975lifetime}, which both addresses the above-listed limitations automatically and allows the use of \textit{multi-period portfolio optimisation} (MPO) to improve portfolio profitability. 
Boyd et al.~\cite{boyd2017multi} demonstrated that MPO was superior to single-period portfolio optimisation, and Li et al.~\cite{li2022multi} built on this work, using multi-period optimisation and an HMM to estimate returns.
In this paper, the HMM of~\cite{li2022multi} will be replaced by our KMRF model, demonstrated in~\cite{pomorski2023improvingportfolio} to be a better regime predictor than an HMM, and it will be shown that this combination allows the construction of long-only, high-profitability portfolios, under a realistic cost model.

\section{Background to this work}\label{sec:background_and_related_work}

\subsection{KAMA+MSR and KMRF models} \label{sec:ch3_and_ch4_models}

This section reviews the two prior models (KAMA+MSR~\cite{pomorski2023improving} and KMRF~\cite{pomorski2023improvingportfolio}) which underpin the current work; the treatment is necessarily brief for reasons of space, and the reader is referred for further details to \cite{pomorski2023improving} and \cite{pomorski2023improvingportfolio}

\subsubsection{KAMA+MSR regime detection model}

The KAMA+MSR model~\cite{pomorski2023improving} combines the strengths of two-state Markov-switching regression (MSR) (a model introduced in~\cite{goldfeld1973markov} and developed further in~\cite{hamilton1989new} and~\cite{krolzig1997markov}) with a technical analysis tool known as \textit{Kaufman's adaptive moving average} (KAMA)~\cite{kaufman1995smarter}. The model overcomes MSR's limitation of being able to detect only two regimes (characterised by high and low volatilities) by adding in KAMA's trend-detection ability, resulting in a model that can effectively detect four regimes (low-variance (LV) bullish, LV bearish, high-variance (HV) bullish, and HV bearish). 
A trading model was proposed in this work which made use of the states of greatest trading value (LV bullish and HV bearish). Results from this were promising, but as a 'nowcasting' model KAMA+MSR inevitably suffered from not being able to take immediate advantage of a change of market regime; it was noted in~\cite{pomorski2023improving} that KAMA+MSR might ideally be used as a regime label generator for machine learning tasks in which the aim is to predict market regimes ex ante.

\subsubsection{KMRF regime prediction model}

The KMRF model~\cite{pomorski2023improvingportfolio} built upon the above-described KAMA+MSR model in using random forest (RF)~\cite{breiman2001random} to now predict, rather than detect, regimes, the predicted regimes being derived from the four KAMA+MSR classes as below:

\begin{itemize}
    \item \textit{Bullish} = LV bullish + extension to the peak of next HV bullish regime.
    \item \textit{Bearish} = HV bearish + extension to the trough of next LV bearish regime.
    \item \textit{Other} = remaining parts of the HV bullish and LV bearish regimes.
\end{itemize}

\noindent The above extensions are carried out in order to avoid losing potential profit through ignoring 'rebound effects' following LV bullish and HV bearish regimes (in the latter case referred to as a 'bear rally').
Additionally, as discussed in ~\cite{pomorski2023improvingportfolio}, it was discovered that the predictions of the KMRF models were best interpreted in a contrarian fashion, with the following strategy adopted:

\begin{itemize}
    \item When a Bullish regime is predicted, a short position should be assumed in the underlying asset.
    \item When a Bearish regime is predicted, a long position should be assumed in the underlying asset.
    \item When the Other regime is predicted, an asset should be sold (if it was bought) or bought back (if it was shorted).
\end{itemize}

\noindent (A contrarian interpretation of the signals will also be used in the current work.) 

The move from regime detection (KAMA+MSR) in~\cite{pomorski2023improving} to prediction (KMRF) in~\cite{pomorski2023improvingportfolio} led to greater profitability. However, it was at the cost of the use of frequent shorting positions, which as noted earlier could be problematic for some investors, an issue the current work will address by showing it is possible to use model predictive control (MPC) to build on these two prior works, allowing also an extension from single-period to multi-period optimisation (MPO), in such a way as to construct a strongly-performing, yet long-only, portfolio.

\subsection{Multi-period optimisation via model predictive control} \label{sec:mpc}

In multi-period optimisation, the portfolio weights are optimised not only for time $t+1$ (single-period optimisation), but also, simultaneously, for times $t+2, t+3, \ldots , t+H$, where $H$ is the prediction horizon, with the optimisation for any $t+n$ building upon the optimisation(s) for the period(s) before. 
In order to implement MPO we will use the mean-variance approach to MPC, a traditional method of selecting portfolio allocations based on a risk-return trade-off which has also recently been found to be robust in a dynamic, multi-period setting~\cite{van2021surprising}. Other MPC variations, based on risk parity or minimum variance, can be used to address the MPO problem, but these are more complex and we have therefore reserved them as topics for future work. 

The construction of the MPC model here will follow that of~\cite{boyd2017multi}. We assume a sequence of allocations 
$\{\textbf{w}_{t+1}, \textbf{w}_{t+2}, \ldots ,\textbf{w}_{t+H}\}$,
executed at time $t$ over a planning horizon $H$, where each allocation (set of portfolio weights) is an $N$-dimensional vector,
where $N$ is the number of assets in the portfolio. 
The objective of MPC with a mean-variance term (namely, the maximisation of returns by the minimisation of risk and costs) can be reduced to the maximisation of the utility function

\begin{equation} \label{eq:mpc}
    \sum_{\tau = t+1}^{t+H}  \hat{\mathbf{r}}_{\tau}^T \mathbf{w}_{\tau} - \gamma^{\textit{sigma}}\mathbf{w}_{\tau}^T \hat{\Sigma}_\tau \mathbf{w}_\tau 
    - \gamma^{trade} \sum_{i=1}^{N} {\hat{TC} (\Delta w_{i,\tau})} -
    \gamma^{hold} \sum_{i=1}^{N} {\hat{HC} (\Delta w_{i,\tau})},
\end{equation}

\noindent for $\tau = t+1, \ldots ,t+H$, where $w_{i,\tau}$ is the weight on asset $i$ at time $\tau$, subject to $w_{i, \tau} \geq 0$, 
$\textbf{1}^T \textbf{w}_{\tau} = 1$, and where $\Delta w_{i,\tau} \triangleq w_{i,\tau} -  w_{i,\tau-1}$.
Additionally, 
$\hat{\textbf{r}}_\tau$ and $\hat{\Sigma}_\tau$ are, respectively, estimates of returns (see Section~\ref{sec:est_returns}) and of the covariance matrix (see Section~\ref{sec:est_covariance__matrix}); $\gamma^{\textit{sigma}}$ is the risk-aversion parameter; $\gamma^{trade}$ is the trading penalty; $\gamma^{hold}$ is the holding penalty; $\hat{TC}$ 
is the estimated transaction cost function (see Section~\ref{sec:est_trading_costs}); and $\hat{HC}$ is the estimated holding cost function, incorporating borrowing fees for shorting assets~\cite{boyd2017multi} (though as this work will construct long-only portfolios the $\hat{HC}$ term will henceforward be dropped).

The risk-aversion parameter $\gamma^{\textit{sigma}}$ controls the trade-off between risk and return; a low $\gamma^{\textit{sigma}}$ corresponds to a high appetite for risk, which can potentially generate high returns during a bullish regime, while a high $\gamma^{\textit{sigma}}$, typically preferred by risk-averse investors, can protect the portfolio during a bearish regime characterised by high volatility. 
The trading penalty controls for turnover. A high $\gamma^{trade}$ leads to a lower turnover, which consumes less profits during rebalancing but exposes the portfolio to holding assets that may not be profitable in the longer term. In contrast, a low $\gamma^{trade}$ increases turnover, thus potentially the costs of trading, but allows the swift allocation of weights from unprofitable to profitable assets. 
The parameters $\gamma^{\textit{sigma}}$ and $\gamma^{trade}$ need to be optimised, in a way which will be described in Sections~\ref{sec:hyperparameter_optimisation} and~\ref{sec:algorithm_implementation}.  

\section{Data and methodology}\label{sec:data_and_methodology}

\subsection{Data used} \label{sec:data}

Two asset classes will be considered, equities (14 assets) and commodities (12 assets). These will be used to create separate regime prediction models for each class, using our KMRF model\cite{pomorski2023improvingportfolio}, these predictions then being used within the MPO model proposed in this paper. It should be noted that while the regime prediction models are built separately by KMRF, the resulting MPO model combines both classes of assets into a mixed portfolio of 26 assets in total. Table~\ref{tab:data_used} lists the component assets for each class, as well as the start dates of the data series, all of which terminate on 29/04/2022. Data up to 30/03/2018 are used for training the KMRF models for the equities and commodities classes, as in~\cite{pomorski2023improvingportfolio}. The period from 02/04/2018 to 29/04/2022, the entirety of which was used as a holdout period in~\cite{pomorski2023improvingportfolio}, is here split into two equal parts, the first for optimisation of the MPO model, the second for testing of the MPO model.
The same range of initial input features as in~\cite{pomorski2023improvingportfolio} were used here, comprising of 
technical (computed using the TA~\footnote{\url{https://github.com/bukosabino/ta}. Last accessed 29 March 2023.} and tsfresh ~\cite{christ2016distributed}) packages), fundamental, and macroeconomic features, before feature selection using the {BorutaShap}~\footnote{\url{https://github.com/Ekeany/Boruta-Shap}. Last accessed 5 April 2023.} package. As in~\cite{pomorski2023improvingportfolio}, this was adapted for time-series use to allow a splitting of validation data using the \textit{purged group time-series split} (PGTS) method, recommended in~\cite{de2018advances}, which respects causality and temporal ordering, and avoids data leakage by leaving a gap between train and validation sets.

\begin{table}[]
\caption{List of asset classes, assets, and start dates.}\label{tab:data_used}
\centering
\begin{tabular}{|l|
>{\columncolor[HTML]{FFFFFF}}l |
>{\columncolor[HTML]{FFFFFF}}l |
>{\columncolor[HTML]{FFFFFF}}l |}
\hline
\cellcolor[HTML]{FFFFFF}\textbf{Asset class}                         & \multicolumn{1}{l|}{\cellcolor[HTML]{FFFFFF}\textbf{Asset}}               & \multicolumn{1}{l|}{\cellcolor[HTML]{FFFFFF}\textbf{Asset label}}    & \textbf{Start date}   \\ \hline
\cellcolor[HTML]{FFFFFF}                                             & S\&P/ASX 200 Index                                                              & ASX                                                                  & 03/05/2001            \\
\cellcolor[HTML]{FFFFFF}                                             & CAC40 Index                                                                     & CAC                                                                  & 29/01/1996            \\
\cellcolor[HTML]{FFFFFF}                                             & DAX Index                                                                       & DAX                                                                  & 29/01/1996            \\
\cellcolor[HTML]{FFFFFF}                                             & FTSE 100 Index                                                                  & FTSE                                                                 & 03/04/2002            \\
\cellcolor[HTML]{FFFFFF}                                             & FTSE MIB Index                                                                  & FTSEMIB                                                              & 06/09/2004            \\
\cellcolor[HTML]{FFFFFF}                                             & KOSPI Index                                                                     & KOSPI                                                                & 24/06/1995            \\
\cellcolor[HTML]{FFFFFF}                                             & MSCI China Index                                                                & MSCI\_China                                                          & 25/12/2002            \\
\cellcolor[HTML]{FFFFFF}                                             & NASDAQ 100 Index                                                                & NASDAQ                                                               & 31/01/2001            \\
\cellcolor[HTML]{FFFFFF}                                             & NIFTY 50 Index                                                                  & NIFTY                                                                & 02/02/2001            \\
\cellcolor[HTML]{FFFFFF}                                             & Nikkei 225 Index                                                                & Nikkei                                                               & 28/02/1995            \\
\cellcolor[HTML]{FFFFFF}                                             & Swiss Market Index                                                              & SMI                                                                  & 31/01/1996            \\
\cellcolor[HTML]{FFFFFF}                                             & S\&P 500 Index                                                                  & SPX                                                                  & 30/01/1995            \\
\cellcolor[HTML]{FFFFFF}                                             & S\&P/TSX Composite Index                                                        & TSX                                                                  & 28/02/1995            \\
\multirow{-14}{*}{\cellcolor[HTML]{FFFFFF}\textbf{Equities}}         & TWSE Index                                                                      & TWSE                                                                 & 30/12/1996            \\ \hline

\cellcolor[HTML]{FFFFFF}                                             & Aluminium Futures                                                         & Aluminium                                                                & 12/08/2004 \\
\cellcolor[HTML]{FFFFFF}                                             & Brent Crude Oil Futures                                                         & Brent                                                                & 12/03/1993 \\
\cellcolor[HTML]{FFFFFF}                                             & Coffee Futures                                                                  & Coffee                                                               & 21/04/1993 \\
\cellcolor[HTML]{FFFFFF}                                             & Copper Futures                                                                  & Copper                                                               & 07/07/1993 \\
\cellcolor[HTML]{FFFFFF}                                             & Corn Futures                                                                    & Corn                                                                 & 23/08/1995 \\
\cellcolor[HTML]{FFFFFF}                                             & Gold Futures                                                                    & Gold                                                                 & 23/11/1993 \\
\cellcolor[HTML]{FFFFFF}                                             & Live Cattle Futures                                                             & Live\_cattle                                                         & 01/10/1992 \\
\cellcolor[HTML]{FFFFFF}                                             & Natural Gas Futures                                                             & Natural\_gas                                                         & 07/05/1993 \\
\cellcolor[HTML]{FFFFFF}                                             & Nickel Futures                                                                  & Nickel                                                               & 10/01/2005 \\
\cellcolor[HTML]{FFFFFF}                                             & Soybeans Futures                                                                & Soybeans                                                             & 17/10/1997 \\
\cellcolor[HTML]{FFFFFF}                                             & Sugar Futures                                                                   & Sugar                                                                & 17/02/1992 \\
\multirow{-14}{*}{\cellcolor[HTML]{FFFFFF}\textbf{Commodities}}         & Wheat Futures                                                                      & Wheat                                                                 & 18/01/1996            \\ \hline

\cellcolor[HTML]{FFFFFF}\textbf{Cash}         & US 3-Month Treasury Bill                                                                      & US3M                                                                 & 18/01/1996            \\ \hline

\end{tabular}
\end{table}

\subsection{Estimation of returns} \label{sec:est_returns}

The KMRF model of~\cite{pomorski2023improvingportfolio} predicts the probability of entering or exiting a specific regime rather than raw asset returns. In order to transform KMRF probabilities into return estimates, we will combine approaches from~\cite{boyd2017multi} and~\cite{li2022multi}, using the following process (in which it should be recalled from Section~\ref{sec:ch3_and_ch4_models} that the KMRF model adopts a contrarian interpretation of trading signals):

\begin{enumerate}
    \item For each asset $i$, following~\cite{boyd2017multi}, calculate a lagged 10-day exponential weighted moving average~\footnote{The exponential weighted moving average (EMA) of a quantity (e.g. a return) is calculated as $EMA_{C_t} = C_t \times (\frac{2}{n+1}) + EMA_{C_{t-1}} \times [1 - (\frac{2}{n+1})]$, where $C_t$ is the current value of the quantity and $n$ is a moving time window, such as 10 days.} ($EMA_{r_{i,t-1}}$) of real daily returns.
    \item For each asset $i$, following~\cite{li2022multi}, but replacing that work's HMM-derived regime predictions by ones from our KMRF~\cite{pomorski2023improving} model, derive an estimated return $\hat{r}_{i,t}$ according to:
    \begin{equation} \label{eq:r-hat}
       \hat{r}_{i,t+1} =  \left\{ \begin{array} {rl} 
       -p_{i,t}^{bull} \times EMA_{r_{i,t}} & \mbox{if prediction is Bullish } \\ 
       p_{i,t}^{bear} \times EMA_{r_{i,t}} & \mbox{if prediction is Bearish} \\ 
       EMA_{r_{i,t}} &  \mbox{if prediction is Other} \end{array} \right.   
    \end{equation}
    \noindent where $p_{i,t}^{bull}$, $p_{i,t}^{bear}$ are the probabilities of asset $i$, at time $t$, entering a Bullish or Bearish (as defined in Section~\ref{sec:ch3_and_ch4_models}) regime, respectively.
\end{enumerate}

Since MPC allocates portfolio weights over a specific time horizon $H$, it requires return estimates at time $t$ for future times $t+1 \dots t+H$; however, KMRF regime predictions are made only for time $t+1$. In order to tackle this problem, a novel two-step solution has been implemented, in which an initial estimate for each required return within the horizon $H$, using available past-time information, is then accuracy-boosted by the use of a Kalman filter:

\begin{itemize}
    \item Each estimated daily return within the horizon $H$ is initially calculated as a lagged return from Eq.~\ref{eq:r-hat}. Specifically, the estimated return $\hat{r}_\tau$, for $\tau = t+1 \dots t+H$, will be calculated using $EMA_{r_{i,\tau - H}}$.
    \item The estimates $\hat{r}_{i,t}$ are then improved by the use of a Kalman filter. Specifically, the accuracy-boosted return estimate $\hat{r}_{i,t}^{KF}$ are calculated as the original return $\hat{r}_{i,t}$ plus the corresponding lagged estimated hidden state (the Kalman filter’s estimate of the difference between the actual return and the estimated return for that asset at a time step $t$). 
\end{itemize}

\subsection{Estimation of the covariance matrix} \label{sec:est_covariance__matrix}

Apart from the estimated returns, the MPC formula (Eq.~\ref{eq:mpc}) requires an estimation of the covariance matrix $\hat{\Sigma}_t$. This work will use the solution suggested by~\cite{boyd2017multi}, namely that $\hat{\Sigma}_t$ should be taken to be a lagged rolling covariance matrix of asset returns over a period of 504 days (two trading years). It will be assumed the estimated risk of underlying assets can be explained by their current risk, the return estimates incorporating information about future risk from the regime prediction, i.e., if Bullish (in our contrarian interpretation, a sell signal) a higher volatility should be expected, while if Bearish (here interpreted as a buy signal) a lower volatility should be expected. (Note that we are here again using the KMRF definitions of 'Bullish' and 'Bearish' given in Section~\ref{sec:ch3_and_ch4_models}.) 

\subsection{Estimation of trading costs} \label{sec:est_trading_costs}

Trading costs will be estimated, following the methods of~\cite{boyd2017multi}, by

\begin{equation} \label{eq:est_tc_model}
   \hat{TC}(\Delta w_{i,t}) = \frac{b}{2} \times P_{i,t}|\Delta w_{i,t}| + EMA_{\sigma_{i,t-1}} \frac{{|\Delta w_{i,t}|}^\frac{3}{2}}
   { {(\frac{EMA_{v_{i,t-1}}}{V_t})}^\frac{1}{2} },
\end{equation}

\noindent where $\Delta w_{i,t}$ is defined as in connection with Eq.~\ref{eq:mpc}, $b$ is the bid-ask spread, $P_{i,t}$ is the price of asset $i$ at time $t$, $EMA_{\sigma_{i, t-1}}$ is the lagged 10-day exponentially weighted moving standard deviation of returns for asset $i$, $EMA_{v_{i,t-1}}$ is the lagged 10-day exponentially weighted moving average of the dollar volume traded for asset $i$, and $V_t$ is the total volume traded by the portfolio at time $t$. In the work of~\cite{boyd2017multi} $b$ was assumed to be only five basis points; however, following our KMRF costs assumptions in~\cite{pomorski2023improvingportfolio}, $b$ will here be set to 0.002 (20 basis points), as this would also include potential explicit costs, such as trading fees. 

\subsection{Hyperparameter optimisation} \label{sec:hyperparameter_optimisation}

Hyperparameter optimisation was used in the construction of both the KMRF and MPC models. For the KMRF model we followed the same process of as in~\cite{pomorski2023improvingportfolio}, using cross-validation (implementing the PGTS method for time-series, recommended in~\cite{de2018advances}) on data up to 30/03/2018, in order to establish optimal settings for the random forest component of KMRF. The first half of the remaining data (02/04/2018--12/03/2020) was then used to tune the MPC parameters $\gamma^{\textit{sigma}}$ and $\gamma^{trade}$, with the data from 27/03/2020--29/04/2022 used to generate the test results in Section~\ref{sec:results}. (A 15-day gap was left before the start of the test period in order to prevent data leakage.) All hyperparameter tuning was carried out using the Optuna~\cite{akiba2019optuna} package, and optimisation, as in the KMRF work~\cite{pomorski2023improvingportfolio}, was carried out with respect to the Sortino ratio,
rather than the Sharpe ratio, as investors are typically less concerned about upside volatility.

\subsection{Algorithm implementation} \label{sec:algorithm_implementation}

\subsubsection{Choice of investment horizon}

In the work of~\cite{boyd2017multi} $H$ was set to 2, while in~\cite{li2022multi} values between 2 and 30 were studied (though it was shown that beyond H=5 the portfolio performance gradually deteriorated). Initial experiments here considered $H = 2, 3, 4, 5$; however, each subsequent value of $H$ resulted in slower calculations, as well as memory issues when $H \geq 4$. We will therefore adopt the value $H=2$, since, in addition to the above considerations, initial experiments showed the $H=2$ portfolio outperformed that for $H=3$. 

\subsubsection{Optimisation of MPC parameters}

This was carried out as described in Section~\ref{sec:hyperparameter_optimisation}, with the search spaces (based on suggestions from~\cite{li2022multi}) $\{0.01, \dots ,1000\}$ for $\gamma^\textit{sigma}$ and $\{0.0001, \dots ,25.0\}$ for $\gamma^\textit{trade}$. The optimal discovered values were $\gamma^\textit{sigma} = 0.1262$ and $\gamma^\textit{trade} = 4.6670$.

\subsubsection{Implementation of MPC}

The MPC component of the KMRF+MPC model was implemented using the the CVXPortfolio~\footnote{\url{https://github.com/cvxgrp/cvxportfolio}. Last accessed 21/02/2023.} package, created by the authors of~\cite{boyd2017multi}.
The package executes the trades on a roll-forward basis. The cost-adjusted returns from these trades are stored in a vector array ready to be used for financial performance measure calculations when the procedure is completed.
Apart from the return and risk estimates ($\hat{r}_{KF_{i,t}}$ and $\hat{\Sigma}_t$), the estimated cost model $\hat{TC}$, and $\gamma^\textit{sigma}$ and $\gamma^\textit{trade}$, CVXPortfolio requires the dates over which the portfolio construction takes place and the initial portfolio value. 
The starting portfolio value is \$26,000, equally distributed among assets (\$1,000 per each asset), except from the cash component which is assumed to be initially \$0. 
As mentioned in Section~\ref{sec:hyperparameter_optimisation}, the dates 02/04/2018--12/03/2020 were used for the optimisation of $\gamma^\textit{sigma}$ and $\gamma^\textit{trade}$, while the period 27/03/2020--29/04/2022 was used for holdout testing, 
with a 15-day gap between the MPC optimisation period and the test period in order to avoid any leakage of the training data to the testing data; the Kalman filter necessary to calculate the accuracy-boosted return estimates $\hat{r}_{i,t}^{KF}$ is recalculated for the holdout sample, to avoid leakage of information between the training and the testing sets. 
It should finally be noted that the MPC algorithm has been restricted to assume the minimum weight of each asset to be 1\% (thus making Eq.~\ref{eq:mpc} subject to $w_{i,\tau} \geq 0.01$ instead of $w_{i,\tau} \geq 0.0$); this is to avoid the optimisation algorithm moving the most of the funds to cash in order to achieve a high Sortino ratio, as the almost zero risk associated with cash could significantly enlarge the Sortino ratio.

\subsection{Selected benchmarks}

Two benchmark portfolios are considered: buy-and-hold and the 1/N (equally-weighted) portfolio. These are commonly used benchmarks both in the industry and in academic finance (for example~\cite{boyd2017multi} used a buy-and-hold benchmark, while~\cite{li2022multi} used 1/N); the buy-and-hold portfolio, due to its minimal costs, can, especially during bull markets, be difficult to beat.

\subsection{Performance metrics}\label{sec:performance_metrics}

We assess the performance of the KMRF-based MPO model according to a set of metrics considered relevant and informative in the industry: \textit{mean excess returns and annualised mean excess returns}, \textit{daily volatility and annualised volatility}, \textit{annualised Sharpe ratio}, \textit{annualised Sortino ratio}, and \textit{maximum drawdown}. We additionally compute the \textit{annualised information ratio} with respect to each of the benchmark portfolios; a value in excess of 1.0 demonstrates the proposed model is more profitable than the benchmark.

\section{Results} \label{sec:results}

\begin{table}[]
\caption{KMRF+MPC portfolio performance vs. benchmarks on the entire holdout set, both with (grey text), and without (regular text) nickel and wheat futures. Note the excess returns are calculated as portfolio cost-adjusted returns minus the risk-free returns of the cash component, namely US three-month treasury bills.}\label{tab:compare_to_benchmarks}
\centering
\begin{tabular}{|
>{\columncolor[HTML]{FFFFFF}}l |
>{\columncolor[HTML]{FFFFFF}}c 
>{\columncolor[HTML]{FFFFFF}}c |
>{\columncolor[HTML]{FFFFFF}}c 
>{\columncolor[HTML]{FFFFFF}}c |
>{\columncolor[HTML]{FFFFFF}}c 
>{\columncolor[HTML]{FFFFFF}}c |}
\hline
\multicolumn{1}{|c|}{\cellcolor[HTML]{FFFFFF}\textbf{}} & \multicolumn{2}{c|}{\cellcolor[HTML]{FFFFFF}\textbf{\begin{tabular}[c]{@{}c@{}}KMRF+MPC\end{tabular}}} & \multicolumn{2}{c|}{\cellcolor[HTML]{FFFFFF}{\textbf{Buy-and-hold}}}                    & \multicolumn{2}{c|}{\cellcolor[HTML]{FFFFFF}{\textbf{1/N}}}                             \\ \hline
Mean excess   returns                                   & \multicolumn{1}{c|}{\cellcolor[HTML]{FFFFFF}{\color[HTML]{606060} 8.20\%}}                & 0.30\%               & \multicolumn{1}{c|}{\cellcolor[HTML]{FFFFFF}{\color[HTML]{606060} 0.23\%}}  & 0.10\%  & \multicolumn{1}{c|}{\cellcolor[HTML]{FFFFFF}{\color[HTML]{606060} 0.13\%}}  & 0.10\%  \\ \hline
{Ann. mean excess returns}                                & \multicolumn{1}{c|}{\cellcolor[HTML]{FFFFFF}{\color[HTML]{606060} {205.41\%}}}              & {74.34\%}              & \multicolumn{1}{c|}{\cellcolor[HTML]{FFFFFF}{\color[HTML]{606060} {65.43\%}}} & {26.21\%} & \multicolumn{1}{c|}{\cellcolor[HTML]{FFFFFF}{\color[HTML]{606060} {31.52\%}}} & {24.21\%} \\ \hline
Volatility                                              & \multicolumn{1}{c|}{\cellcolor[HTML]{FFFFFF}{\color[HTML]{606060} 5.98\%}}                & 2.71\%               & \multicolumn{1}{c|}{\cellcolor[HTML]{FFFFFF}{\color[HTML]{606060} 2.64\%}}  & 0.74\%  & \multicolumn{1}{c|}{\cellcolor[HTML]{FFFFFF}{\color[HTML]{606060} 0.68\%}}  & 0.66\%  \\ \hline
Ann. volatility                                         & \multicolumn{1}{c|}{\cellcolor[HTML]{FFFFFF}{\color[HTML]{606060} 94.93\%}}               & 43.04\%              & \multicolumn{1}{c|}{\cellcolor[HTML]{FFFFFF}{\color[HTML]{606060} 41.89\%}} & 11.79\% & \multicolumn{1}{c|}{\cellcolor[HTML]{FFFFFF}{\color[HTML]{606060} 10.74\%}} & 10.43\% \\ \hline
Ann. Sharpe ratio                                       & \multicolumn{1}{c|}{\cellcolor[HTML]{FFFFFF}{\color[HTML]{606060} 2.16}}                  & 1.73                 & \multicolumn{1}{c|}{\cellcolor[HTML]{FFFFFF}{\color[HTML]{606060} 1.56}}    & 2.22    & \multicolumn{1}{c|}{\cellcolor[HTML]{FFFFFF}{\color[HTML]{606060} 2.93}}    & 2.32    \\ \hline
Ann. Sortino ratio                                      & \multicolumn{1}{c|}{\cellcolor[HTML]{FFFFFF}{\color[HTML]{606060} 5.11}}                  & 2.17                 & \multicolumn{1}{c|}{\cellcolor[HTML]{FFFFFF}{\color[HTML]{606060} 2.67}}    & 2.91    & \multicolumn{1}{c|}{\cellcolor[HTML]{FFFFFF}{\color[HTML]{606060} 3.90}}    & 3.00    \\ \hline
Maximum drawdown                                        & \multicolumn{1}{c|}{\cellcolor[HTML]{FFFFFF}{\color[HTML]{606060} 43.43\%}}               & 28.21\%              & \multicolumn{1}{c|}{\cellcolor[HTML]{FFFFFF}{\color[HTML]{606060} 36.17\%}} & 6.03\%  & \multicolumn{1}{c|}{\cellcolor[HTML]{FFFFFF}{\color[HTML]{606060} 5.19\%}}  & 4.27\%  \\ \hline
Ann. IR (vs. buy-and-hold)                              & \multicolumn{1}{c|}{\cellcolor[HTML]{FFFFFF}{\color[HTML]{606060} {2.11}}}                  & {1.22}                 & \multicolumn{1}{c|}{\cellcolor[HTML]{FFFFFF}--}                             & --      & \multicolumn{1}{c|}{\cellcolor[HTML]{FFFFFF}--}                             & --      \\ \hline
Ann. IR (vs. 1/N)                                       & \multicolumn{1}{c|}{\cellcolor[HTML]{FFFFFF}{\color[HTML]{606060} {1.91}}}                  & {1.23}                 & \multicolumn{1}{c|}{\cellcolor[HTML]{FFFFFF}--}                             & --      & \multicolumn{1}{c|}{\cellcolor[HTML]{FFFFFF}--}                             & --      \\ \hline
\end{tabular}
\end{table}

\begin{figure}
\includegraphics[width=\textwidth]{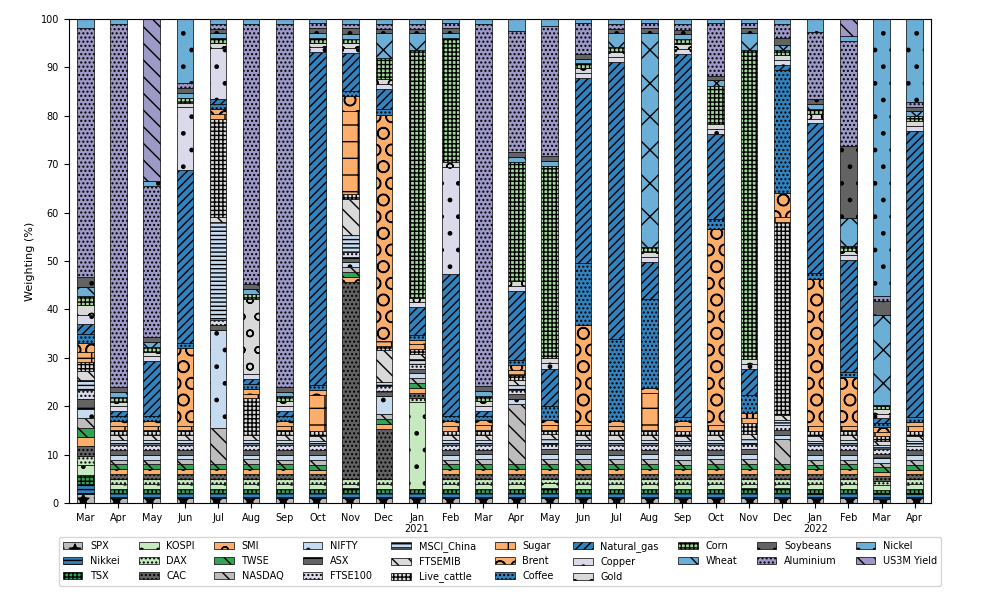}
\caption{KMRF+MPC portfolio monthly asset weights over the test period. Note the very high weighting given to nickel futures (blue dotted column-section) in March 2022, and the increased weighting also of wheat futures.} \label{fig:Portfolio_performance_vs_1N_hold_MPO_1_H_weights}
\end{figure}

\begin{figure}
\includegraphics[width=\textwidth]{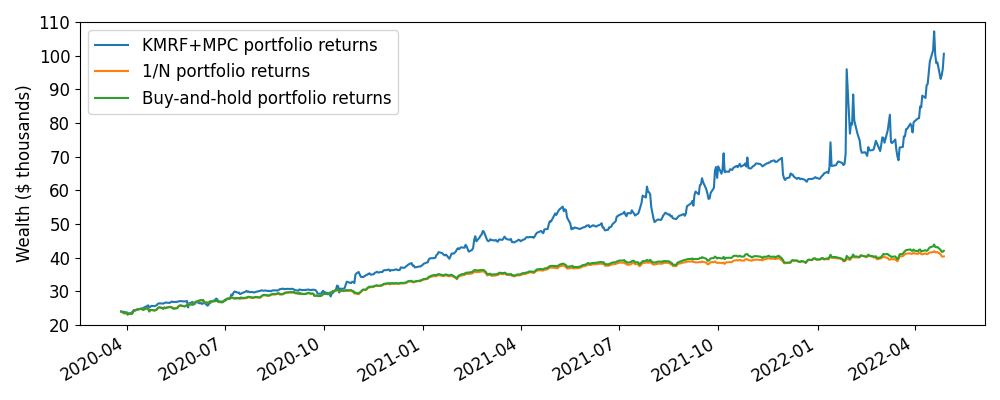}
\caption{KMRF+MPC portfolio test performance vs. its benchmarks, with nickel and wheat excluded from the commodities.} \label{fig:Portfolio_performance_vs_1N_hold_MPO_1_dropped}
\end{figure}

Table~\ref{tab:compare_to_benchmarks} shows the performance of our model compared to the benchmarks, for KMRF+MPC portfolios both with and without nickel and wheat futures, as it can been seen that with these commodities included (grey text in Table~\ref{tab:compare_to_benchmarks}) profits are anomalously high. 
This effect was due to the prices of these two commodities spiking in March 2022 due to the start of war in Ukraine in February 2022. These two assets can also be seen in Fig.~\ref{fig:Portfolio_performance_vs_1N_hold_MPO_1_H_weights} (which shows the weights of assets during each month of the test period) to dominate the KMRF+MPC portfolio during this month. 
Because this situation was exceptional, we decided to primarily compare the models with nickel and wheat futures excluded; Table~\ref{tab:compare_to_benchmarks} and Fig.~\ref{fig:Portfolio_performance_vs_1N_hold_MPO_1_dropped} demonstrate that the KMRF+MPC portfolio continues to substantially outperform its competitors even when the anomalous profits due to the holding of nickel and wheat through March 2022 have been eliminated. 

\begin{figure}
\includegraphics[width=\textwidth]{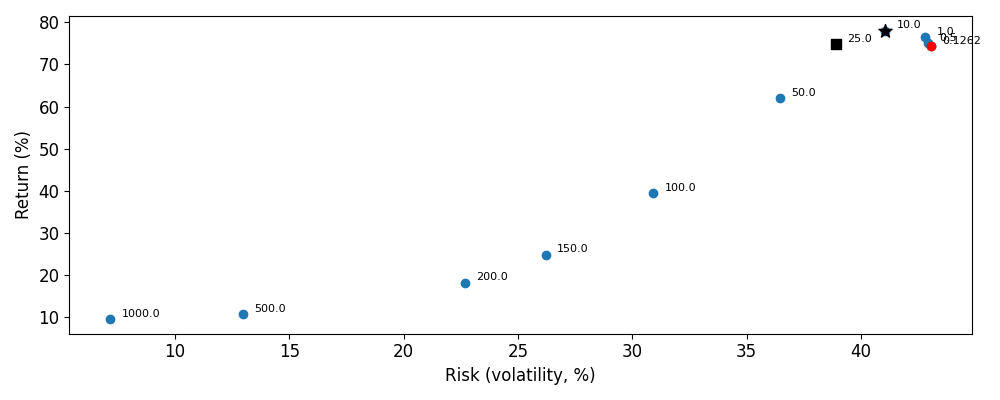}
\caption{Annualised risk-return relationship for a range of values of the risk-aversion parameter $\gamma^{\textit{sigma}}$, calculated for the test data, with the $\gamma^{\textit{sigma}}$ values shown beside each dot. Nickel and wheat are again excluded. The star indicates the $\gamma^{\textit{sigma}}$ with the highest Sortino ratio, and the square that with the highest Sharpe ratio. The red dot indicates the position occupied by the KMRF+MPC portfolio with $\gamma^{\textit{sigma}}= 0.1262$, optimised to deliver the best Sortino ratio during the preceding MPC validation period. } 
\label{fig:risk_return_gamma_sigma}
\end{figure}

Table~\ref{tab:compare_to_benchmarks} and Fig.~\ref{fig:Portfolio_performance_vs_1N_hold_MPO_1_dropped} also evidence, however, that even with the exceptionally high-volatility behaviours of nickel and wheat futures excluded, the profitability of the KMRF+MPC portfolio comes at the cost of substantial volatility, potentially problematic for some investors. However, investors could easily be offered a range of options suitable for differing appetites for risk, by adjusting the MPC risk-aversion ($\gamma ^ \textit{sigma}$) parameter. Fig.~\ref{fig:risk_return_gamma_sigma} shows how $\gamma ^ \textit{sigma}$ affects the test period risk-return tradeoff. It can be seen there is a good potential for return even at a lessened risk level. The $\gamma ^ \textit{sigma}$ choices that would give the highest Sharpe (due to its wide usage) and highest Sortino ratios are shown, as is the position occupied by the KMRF+MPC portfolio during its test period (where it is not guaranteed to deliver the highest Sortino ratio, since its MPC parameters were optimised on a preceding, validation, dataset).

\section{Discussion and conclusions}

This paper built upon a previous work~\cite{pomorski2023improvingportfolio}
in which we proposed a novel regime prediction model, the {KMRF model}, itself built upon a prior work in regime detection, in which we proposed the {KAMA+MSR model}~\cite{pomorski2023improving}, a combination of Kauffman's adaptive moving average and Markov-switching regression. 

In this work we applied multi-period optimisation (MPO) to the predictions of the KMRF model in order to achieve realistic regime-robust asset allocation across multiple assets of two classes (equities and commodities). In order to achieve efficiency and scalability, MPO was here implemented using the method of model predictive control (MPC). By transforming the classification predictions from the KMRF model into return estimates, enhanced by the use of a Kalman filter, and treating the MPC parameters as tunable hyperparameters, we obtained an MPO model, termed here the {KMRF+MPC model}, using which it was possible to obtain an optimal long-only portfolio capable of substantially outperforming its benchmarks (1/N and buy-and-hold portfolios) with respect to the information ratio as well as cumulative return, even after accounting for a more realistically detailed cost model than was used in our KMRF work~\cite{pomorski2023improvingportfolio}.

One potential issue was that while the KMRF+MPC portfolio achieved superior returns to the benchmarks, it could be immediately noticed that this was linked to higher risk. For risk-averse investors this could be problematic, as their investment strategies may simply not allow for higher levels of risk, even if this could lead to substantial profits, though this problem could be addressed either by dropping higher variance assets (e.g. certain commodities or emerging markets equities) from a portfolio, or, as illustrated in Fig.~\ref{fig:risk_return_gamma_sigma}, very simply by adjusting the MPC $\gamma ^ \textit{sigma}$ parameter to balance the risk-return relationship such that it better fits a risk-averse investor’s needs.

It could fairly be said, however, that the KMRF+MPC model presented here is one that appears to thrive on risk. For those investors able to tolerate a high level of risk, rather than taming the model, as above, it may be preferable to instead actively seek out volatile and risky assets for inclusion in KMRF+MPC portfolios, and further work along these lines is ongoing.

\bibliographystyle{unsrt}  
\bibliography{references}

\end{document}